\documentclass[a4paper,12pt]{article}
\usepackage{graphicx}
\usepackage[english]{babel}
\usepackage{amssymb,amsbsy,amsmath,eucal}
\usepackage[compress]{cite}
\usepackage{hyperref}
\hypersetup{
    colorlinks=true,
    linkcolor=blue,
    filecolor=blue,
    urlcolor=blue,
    }
\newcommand{\lab}[1]{\label{#1}}
\newcommand{\re}[1]{(\ref{#1})}
\newcommand{\nn}{\nonumber}
\newcommand{\B}[1]{\boldsymbol{#1}}

\newcommand{\s}[1]{\mathsf{#1}}

\newcommand{\sOm}{\mathsf{\Omega}}
\newcommand{\BOm}{\boldsymbol{\Omega}}
\newcommand{\tdot}{\dddot}
\newcommand{\D}[2]{{\rm d}^{#1}{#2}\,}

\begin{document}

\title{On the free rotation of a polarized spinning-top
 as a test of the correct radiation reaction torque.}

\author{A. Duviryak}
\date{Institute for Condensed Matter Physics of NAS of Ukraine,\\
1 Svientsitskii Street, Lviv, UA-79011, Ukraine \\
              Tel.: +380 322 701496, \
              Fax: +380 322 761158\\
              {duviryak@icmp.lviv.ua}           %  \\
}

\maketitle
\begin{abstract}
The formula for dipole radiation reaction torque acting on a system
of charges, and the Larmor-like formula for the angular momentum
loss by this system, differ in the time derivative term which is the
analogue of the Schott term in the energy loss problem. In the
well-known textbooks this discrepancy is commonly avoided via
neglect of the Schott term, and the Larmor-like formula is
preferred. In the present paper both formulae are used to derive two
different equations of motion of a polarized spinning-top. Both
equations are integrable for the symmetric top and lead to quite
different solutions. That one following from the Larmor-like formula
is physically unplausible, in contrast to another one. This result
is accorded with the reinterpretation of Larmor's formula discussed
recently in the pedagogical literature. It is appeared, besides,
that the Schott  term is of not only academic significance, but it
may determine the behavior of polarized micro- and nanoparticles in
nature or future experiments.\bigskip\\
{\bf Video-abstract:}\\
\url{https://www.dropbox.com/s/im6tmkdrr3oh3ud/EJP_ac578c_Video_Abstr.mov?dl=0}
\end{abstract}


%

%%%%%%%%%%%%%%%%%%%%%%%%%%%% SECTION 1 %%%%%%%%%%%%%%%%%%%%%%%%%%%%%%%%%

\section{Introduction}

It is known that an intensity of the radiation emitted by a system
of charged particles, as it follows from the Larmor formula,
contradicts to the power of the radiation reaction force acting on
the particles, and the difference of these quantities is dependent
of a jerk\footnote{i.e., the time derivative of acceleration
\cite{EPR16}.} and is a rate of the change of so called Schott
energy \cite{Sch12,E-G07}.

The notion of acceleration-dependent Schott energy is known for a
long time, and recently it has attracted attention as a subject of
graduate electrodynamics courses \cite{E-G07,Row10,Gro11,Sin16a}.
The physical interpretation of this notion is still debatable as
referring to ill-defined concepts, such as the point-like charge,
electromagnetic mass etc \cite{Row10,Sin16c}.

Another question is how significant is the Schott term in the energy
balance equation of the system of moving charges. This is a question
of correct application of the Larmor formula and related equations
in practice, in particular, in teaching tasks for students.

In many classic textbooks on electrodynamics, such as by Jackson
\cite[{Sect.\,16.2}]{Jac99}, by Panofsky \& Phillips
\cite[{Sect.\,21-6}]{P-P62}, or by Griffiths
\cite[{Sect.\,11.2.2}]{Gri17}, the Larmor formula is used to derive
the equation of motion of charged particles taking into account the
radiation reaction. One requires, upon derivations, a zero
contribution of the Schott term, at least on average over a time of
interest (in particular, over a period in a periodic motion).

Sometimes there is regarded obvious (as in the Landau \& Lifshitz
textbook \cite[\S\,75]{L-L87E}), that the Schott term is negligibly
small if limited to a nearly stationary motion of a system of
charges\footnote{ The authors of \cite{L-L87E} imply ``a motion which,
although it would be stationary if radiation were neglected,
proceeds with continuum slowing down''. In this clarification the
term `stationary motion'' itself is not wide spread, but in
\cite{L-L87E} it apparently means the bounded (within the
appropriately chosen inertial reference frame) periodic motion or
quasiperiodic motion (for systems of several degrees of freedom; see
\cite[Sect.\,14]{Iro16}).\label{Rem1}}. Therefore, it may seem that the
Larmor formula or its relativistic generalization, the Li\'enard
formula, are sufficient for accounting the effect of radiation
reaction on the motion of charges.

In fact, this is not the case since the value and the role of the
Schott term may be different. This follows from several examples of
the relativistic mechanics of charged particles presented in
literature. For one, in a periodic particle motion, not only an
averaged value \cite{Sin16a} but also the instant value \cite{Nak20}
of the Schott term may occur small. On the other hand, the Li\'enard
formula (where the Schott term is absent) leads to a finite error in
pitch of trajectory when considering a motion of charged particle in
a magnetic field \cite{Sin16b}. Upon uniformly accelerated motion,
the Schott term is increasingly negative \cite{Gro11} while the
radiation reaction in zero. Thus a care must be taken when handling
with the Schott term and Larmor formula.

Fortunately, there are known the equations of motion (and the above
examples are based on them) derived via several ways regardless of
the Larmor formula or energy balance condition
\cite{Sch12,L-L87E,dGS72,Roh90,YaT12}. These are the
Lorentz-Abraham-Dirac equation \cite{Dir38} or its nonrelativistic
predecessor known as the Abraham-Lorentz equation \cite{Lor09} which
both are widely  accepted.

In turn, it is possible from these equations to deduce
unambiguously the energy balance equation \cite[\S\,75]{L-L87E}
containing the Schott term, whatever its magnitude.

Similarly to the intensity of radiation, but less often, there is
considered in textbooks the flux of angular momentum which being
lost by charges via radiation.

For a single particle in a conservative central field the loss rate
of the angular momentum is given in the textbook \cite{Jac99}; for a
system of charges the corresponding formula was derived in
\cite[\S\,72,\S\,75]{L-L87E}.

As in the case of energy, this formula does not agree with the
torque of radiation reaction forces against charges, and the
difference is equal to the rate of change of a vector quantity
referred here to as the {\em Schott angular momentum}. The
aforementioned textbooks suggest, by analogy with the case of
energy, to neglect the corresponding Schott term in the balance
equation of angular momentum as well.

It is demonstrated in the present work that, in general, this
suggestion is not correct and may lead to a mistaken picture of
behavior of even nonrelativistic systems. For this purpose we
consider a free spinning-top possessing a
constant proper electric dipole moment.

The translational motion of such a spinning-top is described by a
physical solution of the Abraham-Lorentz equation for a free
particle, i.e., the trivial motion by inertia. In order to include
the rotational degrees of freedom, the Abraham-Lorentz equation is
not sufficient: one have apply the balance equation of angular
momentum. The question arises: should a Schott term be included in
this equation? In both cases (with and without this term) the
balance equation can be reduced to nonlinear equations of the Euler
type which are integrable provided the spinning-top is axially
symmetric. These equations involve higher-order kinematics and
themselves are of pedagogical interest as new solvable examples of
rigid body dynamics. The solutions found in both cases differ
essentially from each other, and represent completely different
evolutions of the spinning-top. Thus, there is a question of
choosing correct angular momentum balance equation, which is
discussed in final section.

%---------- To Rem 1 of Ref 1 -------------------------------
The considered problem is not purely academic. Nowadays,
nanoparticles in optical traps can be spun up to GHz
\cite{RDHD18,AXBJGL20}. On the other hand, the dipole moment of some
artificially created nanocrystals reaches hundreds and thousands
Debyes \cite{S-K06,F-PBL14} (and moreover expected up to
10$^7\,$ D \cite{MNS03}). Under such trends, the effects of
radiative spindown may soon become observable experimentally
\cite{Duv20a}, and their theoretical description should be relevant.

%%%%%%%%%%%%%%%%%%%%%%%%%%%% SECTION 2 %%%%%%%%%%%%%%%%%%%%%%%%%%%%%%%%%

\section{Angular momentum balance equation for a system of charges.}
\renewcommand{\theequation}{2.\arabic{equation}}
\setcounter{equation}{0}

Let us consider a nonrelativistic system of charges $q$ with masses
$m$ situated in positions $\B r(t)$ and moving with velocities
$\dot{\B r}\equiv \D{}{\B r}/\D{}t$ much smaller than speed of light
$c$. Such a system loses an energy via the dipole radiation (other
multipole components are negligibly small in the nonrelativistic
approximation). Similarly, the system loses an angular momentum, and
this loss can be taken into account variously. The Landau \&
Lifshitz textbook suggests two methods.

The 1st method, similar to deriving the Larmor formula, consists in
accounting the flux of the angular momentum of the dipole radiation
through the sphere embracing the charges, and leads to the following
formula (see \cite{L-L87E}, \S 72, Problem 2, equation (3); also
equation (75.7))
%
%           Equation 2.1
\begin{equation}\lab{2.1}
\frac{\D{}{\B L}}{\D{}t}=-\frac2{3c^3}\,\dot{\B{\frak
d}}\times\ddot{\B{\frak d}}
\end{equation}
for the angular momentum $\B L=\sum m\,\B r\times\B v$, where
$\B{\frak d}=\sum q\B r$ is a dipole moment of a system; here the
summation runs over all particles of the system. The 2nd method is
based on calculating the torque of the Abraham-Lorentz forces of a
radiation reaction; it yields another formula (look for unnumbered
equation in \cite[\S\,75]{L-L87E}):
%
%           Equation 2.2
\begin{equation}\lab{2.2}
\frac{\D{}{\B L}}{\D{}t}=\frac2{3c^3}\,\B{\frak
d}\times\tdot{\B{\frak d}}.
\end{equation}
To agree with the formula \re{2.1}, the authors of \cite{L-L87E}
represented the expression on the right-hand side (r.-h.s.) of
\re{2.2} as follows:
%
%           Equation 2.3
\begin{equation}\lab{2.3}
\B{\frak d}\times\tdot{\B{\frak d}}=\frac{\D{}{}}{\D{}{t}}\B{\frak
d}\times\ddot{\B{\frak d}}-\dot{\B{\frak d}}\times\ddot{\B{\frak d}},
\end{equation}
and then claimed that total time derivative (the 1st term in
r.-h.s.) vanish upon averaging over a stationary motion (implying a
nearly stationary motion; see footnote$^{\ref{Rem1}}$ on page
\pageref{Rem1}). The result is the formula \re{2.1}.

It is convenient, by analogy to the notion of Schott energy, to introduce
the vector quantity referred here to as the {\em Schott angular momentum}:
%
%           Equation 2.4
\begin{equation}\lab{2.4}
\B L_{\rm S}\equiv\frac{2}{3c^3}\B{\frak d}\times\ddot{\B{\frak
d}}=\frac{2}{3c^3}\frac{\D{}{}}{\D{}t}\B{\frak d}\times\dot{\B{\frak
d}}.
\end{equation}
Then the r.-h.s. of \re{2.2} can be reduced, by accounting \re{2.3},
to the r.-h.s. of \re{2.1} plus the {\em Schott term} $\dot{\B
L}_{\rm S}$ (i.e., the rate of change of the Schott angular
momentum) which is considered in \cite[\S\,75]{L-L87E} negligibly
small.

In Jackson textbook \cite[{Sect.\,16.2}]{Jac99} the Abraham-Lorentz
equation for a particle in the external central field $V(r)$,
%
%           Equation 2.5
\begin{equation}\lab{2.5}
m\dot{\B v}=\B F_{\!\!\rm ex}+\frac{2q^2}{3c^3}\ddot{\B
v},\qquad\mbox{where}\quad \B F_{\!\!\rm
ex}=-\frac{\D{}{V}}{\D{}r}\frac{\B r}{r},
\end{equation}
is used to derive the balance equation of the particle angular momentum
$\B L=m\B r\times{\B v}$:
%
%           Equation 2.6
\begin{equation}\lab{2.6}
\frac{\D{}{\B L}}{\D{}t}=\frac{2q^2}{3c^3}\B r\times\ddot{\B v}.
\end{equation}
Here  is used the fact that the torque of an external central force
vanish, $\B r\times\B F_{\!\!\rm ex}=0$. The remaining term in
r.-h.s. can again be presented as follows:
%
%           Equation 2.7
\begin{equation}\lab{2.7}
\frac{2q^2}{3c^3}\B r\times\ddot{\B v}=\tau_0\frac{\D{2}{\B
L}}{\D{}{t^2}}-\frac{2q^2}{3c^3}\B v\times\dot{\B v},
\end{equation}
where $\tau_0=2q^2/(3mc^3)$ is a small parameter of the dimension of
time (for the electron we have
$\tau_0\approx(2/3){\times}10^{-23}$s.). One supposes in
\cite[{Sect.\,16.2}]{Jac99} that the angular momentum $\B L$ little
changes during the time $\tau_0$. Thus the 1st  term (i.e., the
Schott term) $\dot{\B L}_{\rm S}=\tau_0\ddot{\B L}$ can be
neglected, and the equation \re{2.6} reduces to the form:
%
%           Equation 2.8
\begin{equation}\lab{2.8}
\frac{\D{}{\B L}}{\D{}t}=-\frac{2q^2}{3c^3}\B v\times\dot{\B v}.
\end{equation}
Then, the acceleration in the 2nd term of r.-h.s. is to be expressed
via the external force by an unperturbed equation of motion
\footnote{Let us note that in \cite[\S\,75]{L-L87E} this method is
applied not only to balance equations but it is also used for the
reduction of accelerations and jerks in r.-h.s. of Abraham-Lorentz
\re{2.5} and Lorentz-Abraham-Dirac equations. The reduced in such a
way equations are known in literature as Landau-Lifshitz equations.
\lab{Rem2}}. Finally, one obtains the balance equation:
%
%           Equation 2.9
\begin{equation}\lab{2.9}
\frac{\D{}{\B
L}}{\D{}t}=-\frac{\tau_0}{m}\left\langle\frac1r\frac{\D{}{V}}{\D{}t}
\right\rangle\B L,
\end{equation}
in which the averaging $\langle\dots\rangle$ is performed over a
particle orbit\footnote{Averaging is not necessary, but simplifying.
The exact solution of the equation \re{2.9} for the case of Coulomb
potential, but without averaging, was found in literature later
\cite{Raj08}.}. It follows from this equation that $|\dot{\B
L}|=O(\tau_0)$, then $|\dot{\B L}_{\rm S}|=|\tau_0\ddot{\B
L}|=O(\tau_0^3)$, which is negligibly small as compared to r.-h.s.
of \re{2.9}. Thus the Schott term was neglected reasonably.

Before giving a counterexample in which the neglect of
the Schott term is erroneous, the following caveat should be made.

The equations presented in this section reveal the inconsistency
that they are based on the Newtonian (i.e., nonrelativistic)
dynamics, supplemented by the dipole radiation correction.
The latter arises from the relativistic theory as a term of order $1/c^3$
in the  $1/c$ expansion of the Li\`enard-Wiechert potentials.
Other multipole contributions have orders of magnitude $1/c^5$ and
higher, and are negligibly small for slow-moving charges.

Along with dipole radiation terms, the dynamics should take into
account post-Newtonian corrections of the order of $1/c^2$, which
arise from relativistic kinematics and potentials of interparticle
interaction \cite[\S\,65]{L-L87E}.
However, for a conservative system (such as a closed system of
charges or a particle in the central field), post-Newtonian
corrections are also conservative, and lead to effects other than
the dissipative radiative damping.
For example, in the classical hydrogen atom problem, the
post-Newtonian terms does not destroy the periodicity of motion, but
only cause a perihelion advance, while the radiation reaction
leads to a fall to the center. (In the quantum problem, the
$c^{-2}$--corrections split the spectral lines, while $c^{-3}$ broad them).
The nonrelativistic approximation is used in other cases, for
example, when taking into account the radiation effects in the
Penning trap \cite[Sect.\,7.4]{Vog18}; the post-Newtonian effects are analyzed
separately \cite[Sect.\,7.6]{Vog18}; see also \cite{YPM15}.

The next sections will study the effect of radial reaction on the
rotational motion of a composite particle considered as a
nonrelativistic rigid body%
%----------------------------To Rem 2 of Ref 1-------------------------
\footnote{A modern, rather simple and exhaustive for the reader exposition of
the rigid body dynamics can be found in the textbook \cite[Sect.\,11]{Iro16}.}
%-------------------------------------------------------------------
with a proper dipole moment. From a practical viewpoint, such a
consideration is appropriate, since in the mentioned in the
Introduction examples \cite{RDHD18,AXBJGL20} of a 100 nm particle
spined up to GHz, the velocity of its components does not exceed
10$^{-6}c$. At this point, however, the principal inconsistency
arises as the notion of rigid body contradicts to the relativity
\cite[Sect.\,45]{Pau58}. Instead, one should consider a (confined)
deformed elastic medium which, in the post-Newtonian approximation,
can be close to the rigid one, the so called quasirigid body
\cite{XTW04}. Mechanical properties of this body are determined by
the energy-momentum tensor of the medium, and can be defined in such
a way that, in the post-Newtonian approximation, they entirely are
expressed via corrections to the inertia tensor.

Therefore, at least within this model, $c^{-3}$ radiation effects can be
combined directly with the nonrelativistic rigid body dynamics. This
consideration will be used in subsequent sections.

In general case of quasirigid body, the post-Newtonian $c^{-2}$
corrections to the dynamics are more complicated but still
conservative, and their neglect should not misrepresent the dissipative
radiation effects.

%%%%%%%%%%%%%%%%%%%%%%%%%%%% SECTION 3 %%%%%%%%%%%%%%%%%%%%%%%%%%%%%%%%%

\section{Equations of motion of free polarized spinning-top}
\renewcommand{\theequation}{3.\arabic{equation}}
\setcounter{equation}{0}

Let a system of charges be a composite particle considered as a
rigid body, i.e., a spinning-top.
If the spinning-top is free, then its translational motion is a
trivial motion by inertia. A rotational motion of the spinning-top
can be described by means of the angular momentum balance equation.
Let us represent for this purpose an arbitrary point $\B r(t)$ of
the top as follows: $\B r(t)=\s O(t)\B\rho$, where $\s O(t)\in\
$SO(3) is a rotation matrix, and $\B\rho$ is constant (in time)
position of this point in the proper reference frame of the top.
Hence the relations follow:
%
%           Equation 3.1
\begin{eqnarray}\lab{3.1}
\B v&\equiv&\dot{\B r}=\dot{\s O}\B\rho=\s O(\BOm\times\B\rho),\nn\\
\dot{\B v}&=&\s
O\{\BOm\times(\BOm\times\B\rho)+\dot{\BOm}\times\B\rho\},
\end{eqnarray}
where $\BOm$ is the angular velocity vector of the spinning-top in
its proper reference frame which is dual to the skew-symmetric matrix
$\sOm\equiv\s O^{\rm T}\dot{\s O}$. In general, an arbitrary vector
$\B\alpha$ in the proper reference frame and its image $\B a=\s O\B\alpha$
in the laboratory reference frame are related as follows:
%
%           Equation 3.2
\begin{equation}\lab{3.2}
\dot{\B a}=\s O\{\dot{\B\alpha}+\BOm\times\B\alpha\}.
\end{equation}
These kinematic relations can be used in the angular
momentum balance equation in order to derive the Euler type equation of a
rotational motion.

At this point the the dilemma arises -- which balance equation to
choose: \re{2.1} or \re{2.2} ? Let us consider both possibilities.

Substituting the relations \re{3.1} into \re{2.1} yields the
equation:
%
%           Equation 3.3
\begin{eqnarray}\lab{3.3}
\s I\dot{\BOm}+\BOm\times\s I\BOm &=&-\frac2{3c^3}\{(\B
d\times\BOm)^2\BOm+(\B d\cdot(\BOm\times\dot{\BOm}))\B d\},
\end{eqnarray}
where $\s I=||I_{ij}||$ ($i,j=1,2,3$) is the inertia tensor, and $\B
d\equiv\s O^{\rm T}\B{\frak d}=\sum m\B\rho$ is a constant dipole
moment of the spinning-top in its proper reference frame.

Similarly, the usage of \re{3.1} and \re{3.2} in \re{2.2} leads to the equation:
%
%           Equation 3.4
\begin{eqnarray}\lab{3.4}
\s I\dot{\BOm}+\BOm\times\s I\BOm &=&\frac2{3c^3}\B d\times\{\B
d\times(\Omega^2\BOm-\ddot{\BOm})\nn\\
&&\hspace{10ex}{}+(\B d\cdot\dot{\BOm})\BOm+2(\B d\cdot\BOm)\dot{\BOm}\};
\end{eqnarray}
where $\Omega\equiv|\BOm|$.
This equation is a counterpart of the Abraham-Lorentz equation
for the spinning-top, and it will be referred here to as the
{\em Abraham-Lorentz-Euler (ALE) equation}. The equation \re{3.3}
with the Schott term neglected (as compared to \re{3.4})
will be called as the {\em truncated ALE equation}.

In order to have a complete description of the spinning-top
dynamics in space we complement the Euler type equations \re{3.3} or
\re{3.4} by the Poisson equations:
%
%           Equation 3.5
\begin{eqnarray}\lab{3.5}
\Omega_1 &=& \dot\varphi\sin\theta\sin\psi +
\dot\theta\cos\psi,\nn\\
\Omega_2 &=& \dot\varphi\sin\theta\cos\psi -
\dot\theta\sin\psi,\nn\\
\Omega_3 &=& \dot\varphi\cos\theta + \dot\psi,
\end{eqnarray}
relating the components of the angular velocity
$\BOm=\{\Omega_1,\Omega_2,\Omega_3\}$ with the Euler angles
$\varphi, \theta, \psi$.

%%%%%%%%%%%%%%%%%%%%%%%%%%%% SECTION 4 %%%%%%%%%%%%%%%%%%%%%%%%%%%%%%%%%

\section{The dynamics of axially-symmetric spinning-top}
\renewcommand{\theequation}{4.\arabic{equation}}
\setcounter{equation}{0}

\subsection{Truncated Abraham-Lorentz-Euler equations}

Henceforth the specific case will be considered of the axially-symmetric
spinning-top with a dipole moment directed along the symmetry axis:
%
%           Equation 4.1
\begin{equation}\lab{4.1}
I_{ij}=I_i\delta_{ij}\quad (i,j=1,2,3), \quad I_2=I_1;\qquad
d_1=d_2=0,\quad d_3\equiv d
\end{equation}
(there is no summation over $i$). Besides, we assume $I_1\ne0$,
$I_3\ne0$. Then the equation \re{3.3} splits into the following
nonlinear set:
%
%           Equation 4.2-4
\begin{eqnarray}
\dot\Omega_1-(1-I_3/I_1)\Omega_2\Omega_3=-\tau\Omega_\bot^2\Omega_1,
\lab{4.2}\\
\dot\Omega_2+(1-I_3/I_1)\Omega_1\Omega_3=-\tau\Omega_\bot^2\Omega_2,
\lab{4.3}\\
\dot\Omega_3=-\tau\Omega_\bot^2\Omega_3; \lab{4.4}
\end{eqnarray}
here $\Omega_\bot=|\BOm_\bot|=\sqrt{\Omega_1^2+\Omega_2^2}$, where
$\BOm_\bot=\{\Omega_1,\Omega_2,0\}$, and
again the small parameter of the dimension of time
is introduced: $\tau=2d^2/(3I_1c^3)$.

In order to solve this set of differential equations let us first
multiply \re{4.2} by $\Omega_2$, \re{4.3} by $\Omega_1$, and add them yielding
the equation for $\Omega_\bot$:
%
%           Equation 4.5
\begin{equation}\lab{4.5}
\dot\Omega_\bot=-\tau\Omega_\bot^3.
\end{equation}
The solution of this equation, $\Omega_\bot(t)=\Omega_{\bot0}/R(t)$,
where
%
%           Equation 4.6
\begin{equation}\lab{4.6}
R(t)\equiv\sqrt{1+2\tau\Omega_{\bot0}^2t}, \quad
\Omega_{\bot0}\equiv\Omega_{\bot}(0),
\end{equation}
is to be substituted into \re{4.4}, and then into \re{4.2}, \re{4.3}
reducing a whole set of equations to a linear one. In view of an axial
symmetry of the spinning-top it is convenient to choose the initial
data in such a way that $\Omega_{10}\equiv\Omega_{1}(0)=0$. Hence
$\Omega_{\bot0}=|\Omega_{20}|$, and a solution takes the form:
%
%           Equation 4.7
\begin{eqnarray}
\Omega_1(t)&=&\frac{\Omega_{20}}{R(t)}\sin\!\left\{\frac{\tilde\Omega_{30}[R(t){-}1]}{\tau\Omega_{20}^2}\right\},\quad
\Omega_2(t)=\frac{\Omega_{20}}{R(t)}\cos\!\left\{\frac{\tilde\Omega_{30}[R(t){-}1]}{\tau\Omega_{20}^2}\right\},
\nn\\
\Omega_3(t)&=&\Omega_{30}/R(t), \lab{4.7}
\end{eqnarray}
where $\Omega_{30}\equiv\Omega_{3}(0)$, and
$\tilde\Omega_{30}\equiv(1-I_3/I_1)\Omega_{30}$.

With this solution, one can determine the orientation of the
spinning-top in space at any time. Let us invert for this purpose
the Poisson equations with respect to time derivatives of Euler
angles, so reducing the set \re{3.5} to a normal form:
%
%           Equation 4.8
\begin{eqnarray}
\dot\varphi &=& (\Omega_1\sin\psi + \Omega_2\cos\psi)/\sin\theta,
\nn\\
\dot\theta &=& \Omega_1\cos\psi - \Omega_2\sin\psi, \nn\\
\dot\psi  &=& \Omega_3 - (\Omega_1\sin\psi +
\Omega_2\cos\psi)\cot\theta. \lab{4.8}
\end{eqnarray}
Substituting the solution \re{4.6}-\re{4.7} into r.-h.s. of \re{4.8}
and changing the time variable and one of angle variables:
%
%           Equation 4.9
\begin{equation}\lab{4.9}
t\mapsto\vartheta=[R(t){-}1]/(\tau\Omega_{\bot0}^2), \quad
\psi\mapsto\tilde\psi=\psi-\tilde\Omega_{30}\vartheta
\end{equation}
simplifies the set \re{4.8} to the form:
%
%           Equation 4.10
\begin{eqnarray}
\D{}\varphi/\D{}\vartheta&=&\Omega_{20}\cos\tilde\psi/\sin\theta
\nn\\
\D{}\theta/\D{}\vartheta&=&-\Omega_{20}\sin\tilde\psi, \nn\\
\D{}{\tilde\psi}/\D{}\vartheta&=&\Omega_{30}I_3/I_1 -
\Omega_{20}\cos\tilde\psi\cot\theta. \lab{4.10}
\end{eqnarray}
All initial orientations of the free spinning-top in an isotropic
space are physically equivalent. Thus it is sufficient to find any
solution of the equations \re{4.10}, for example, the following one:
%
%           Equation 4.11
\begin{eqnarray}
&&\tilde\psi=0 \quad\Longrightarrow\quad
\psi=\tilde\Omega_{30}\vartheta, \nn\\
&&\theta=\arctan(\Omega_{20}/\tilde\Omega_{30}), \nn\\
&&\varphi=\frac{\Omega_{20}}{\sin\theta}\vartheta.\lab{4.11}
\end{eqnarray}
This solution coincides, up to the substitution $t\mapsto\vartheta$,
with the solution of the unperturbed Euler-Poisson equations for the free
symmetric spinning-top, i.e., of the equations \re{4.8} with
the functions $\Omega_i(t)$ ($i=1,2,3$) to be solutions of the Euler
equations \re{4.2}-\re{4.4} with zero r.-h.s.

This means that the spinning-top precesses with a constant
inclination angle $\theta$ (i.e., $\dot\theta=0$), and the rates of
precession $\dot\varphi$ and proper rotation $\dot\psi$ decrease in
time as $1/R(t)\sim 1/\sqrt{t}$, i.e, they go to zero in the limit
$t\to\infty$.

\subsection{Reduced Abraham-Lorentz-Euler equations}

Let us consider now the Abraham-Lorentz-Euler equation \re{3.4}. It
contains in r.-h.s. the 2nd derivative $\ddot{\BOm}$ multiplied by a
small parameter $d^2/c^3\propto \tau$, i.e., this equation is a
singularly perturbed one, and by this peculiarity is similar to the
Abraham-Lorentz equation \re{2.5}. Such equations possess redundant
solutions which are non-analytic in a perturbation parameter $\tau$
and describe non-physical runaway motion. The problem can be removed
by reducing higher-order derivatives in small perturbation terms by
usage of unperturbed equations of motion and their differential
consequences. This procedure yields physically admissible equations
of motion of Landau-Lifshitz type (see footnote$^{\ref{Rem2}}$ on
page \pageref{Rem2}).

Splitting by components the equation \re{3.4} for the axially
sysmmetric spinning-top \re{4.1} and taking in r.-h.s. into account
the unperturbed Euler equations (i.e., equations \re{4.2}-\re{4.4}
with zeros in r.-h.s.) together with their differential consequences
yields the following set of {\em reduced Abraham-Lorentz-Euler
equations} \cite{Duv20a}:
%
%           Equation 4.12-14
\begin{eqnarray}
\dot\Omega_1-\Omega_2\tilde\Omega_3=-\tau\{\Omega_\bot^2+(I_3/I_1)^2\Omega_3^2\}\Omega_1,
\lab{4.12}\\
\dot\Omega_2+\Omega_1\tilde\Omega_3=-\tau\{\Omega_\bot^2+(I_3/I_1)^2\Omega_3^2\}\Omega_2,
\lab{4.13}\\
\dot\Omega_3=0, \lab{4.14}
\end{eqnarray}
where $\tilde\Omega_3\equiv(1-I_3/I_1)\Omega_3$. It follows from \re{4.14}
that $\Omega_3=\,$const. Other equations  \re{4.12}, \re{4.13} can be
integrated out similarly to the equations \re{4.2}, \re{4.3}.
Using notations of subsection 4.1 and the same initial condition
$\Omega_{10}=0$, one obtains the solution:
%
%           Equation 4.15-16
\begin{eqnarray}
&&\Omega_1=\Omega_{20}\Phi(t)\,\sin{\tilde\Omega_3t},\quad\Omega_2=\Omega_{20}\Phi(t)\,\cos{\tilde\Omega_3t},
\lab{4.15}\\
&& \Phi(t)\equiv\sqrt{\frac{\beta-1}{\beta{\rm e}^{\,2t/\sigma}-1}},\quad
\beta\equiv1+\frac{I_3^2\Omega_3^2}{I_1^2\Omega_{20}^2},\quad \sigma\equiv\frac{I_1^2}{I_3^2\Omega_3^2\tau}.  \lab{4.16}
\end{eqnarray}
In the limit $\Omega_3\to0$ the function $\Phi(t)$ reduces to $1/R(t)$;
see \re{4.6}.

The Poisson equations \re{4.8} for this case take the form:
%
%           Equation 4.17
\begin{eqnarray}
\dot\varphi&=&\Omega_{20}\Phi(t)\cos\bar\psi/\sin\theta,
\nn\\
\dot\theta&=&-\Omega_{20}\Phi(t)\sin\bar\psi, \nn\\
\dot{\bar\psi}&=&\Omega_{30}I_3/I_1 -
\Omega_{20}\Phi(t)\cos\bar\psi\cot\theta, \lab{4.17}
\end{eqnarray}
where $\bar\psi=\psi-\tilde\Omega_3t$.

Again, in view of the space isotropy it is sufficient to have any
particular solution of the equations \re{4.17}. It can be found
numerically since a search of analytical solution failed. Instead,
one can directly examine the existence of the particular solution
with the following asymptotic behavior at $t\to\infty$:
%
%           Equation 4.18
\begin{eqnarray}
\varphi\sim (I_3/I_1)\Omega_3t,\quad \psi\sim\tilde\Omega_3t,\quad
\theta\sim {\rm e}^{-t/\sigma}.\lab{4.18}
\end{eqnarray}
Therefore, in the limit $t\to\infty$ the spinning-top tends to a
vertical position ($\theta\to0$) in which it will rotate with the
proper angular velocity $\dot\phi+\dot\psi=\Omega_3$. This picture
does not agree with that one following from the truncated ALE
equations and stating that the radiation reaction torque should
reduce at $t\to\infty$ any rotary motion to zero, in accordance with
the solution \re{4.11}.

%%%%%%%%%%%%%%%%%%%%%%%%%%%% SECTION 5 %%%%%%%%%%%%%%%%%%%%%%%%%%%%%%%%%

\section{Discussion}
\renewcommand{\theequation}{5.\arabic{equation}}
\setcounter{equation}{0}

Angular momentum balance equation for a system of charges without
the Schott term \re{2.1} and with it \re{2.2} lead to different
rotational evolutions of the free symmetric polarized spinning-top.

In the first case, the spinning-top moves in the same way as the
free Euler spinning top, however, slowing down in the asymptotics
$t\to\infty$ all angular velocities $\dot\phi$, $\dot\psi$ (and
$\dot\theta=0$) in proportion to the power law $\sim1/\sqrt{t} $;
see figure 1. It is strange that the speed of proper rotation
$\dot\psi$ decreases to zero. Indeed, one can imagine an equivalent
spinning-top, i.e., with the same inertia tensor and dipole moment,
in which all charges are located on the symmetry axis. In this case
no charges rotate around this axis. Then where does the braking
torque relative to this axis come from ?%
%-------------------------- Correction 04.02.2022 -----------------
\footnote{This argument is valid only in the framework of the
dipole approximation adopted in this paper. The braking torque can
also come from the neglected here higher-order multipole
contributions of a radiation generated by charges rotating around
the symmetry axis. In general, they cause slowdown and, ultimately,
stop the proper rotation of the spinning top during a time much
longer than the time frame for the motion discussed in this paper.
\label{Rem3}}
%-----------------------------------------------------------------

In the second case, the spinning-top reduces exponentially the
inclination angle and stabilizes asymptotically its orientation and
proper rotation with the angular velocity $\Omega_3$, angular
momentum $L_3=I_3\Omega_3$ and the energy $E=I_3\Omega_3^2/2$; see
figure 2. This behavior seems more plausible than that of the first
case (but see again the footnote$^{\ref{Rem3}}$). Nevertheless,
stronger arguments in favor of one or another equation are
necessary.

%---------------------- Figure 1-----------------------------
\begin{figure}[p]
\begin{center}
\rule{4ex}{0ex}\includegraphics[scale=0.8]{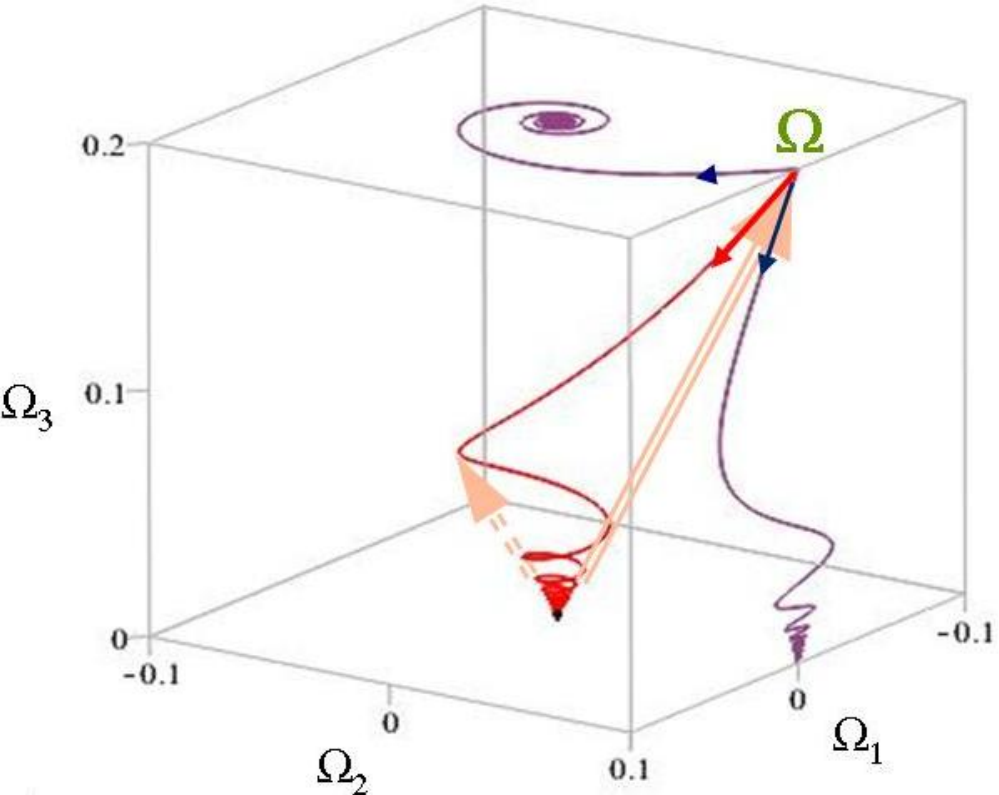}\hspace{3ex}
\includegraphics[scale=0.7]{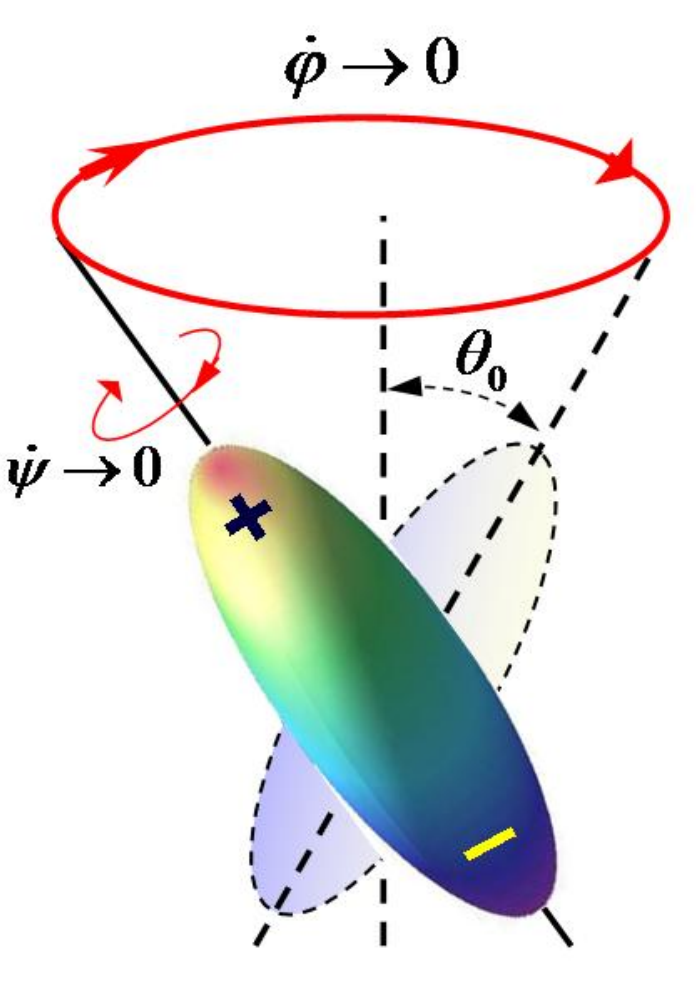}
\vspace{-2ex} \caption{Left -- hodograph and its projections onto
coordinate planes of angular velocity $\BOm$ in the proper reference
frame of the axially-symmetric polarized spinning-top according to
the truncated ALE equations; the rate of principal moments of
inertia: $I_3/I_1=2/5$; initial angular velocity:
$\BOm_0=\{0,0.1,0.2\}$ in $1/\tau$ units. Right -- qualitative
behavior of the spinning-top in space: inclination angle is constant
$\theta_0$, and speeds of precession $\dot\varphi$ and proper
rotation $\dot\psi$ decrease in asymptotics $t\to\infty$ to zero as
$\sim 1/\sqrt{t}$. \vspace{2ex}}
%\end{center}
%\end{figure}
%\begin{figure}[hb]
%\begin{center}
\includegraphics[scale=0.7]{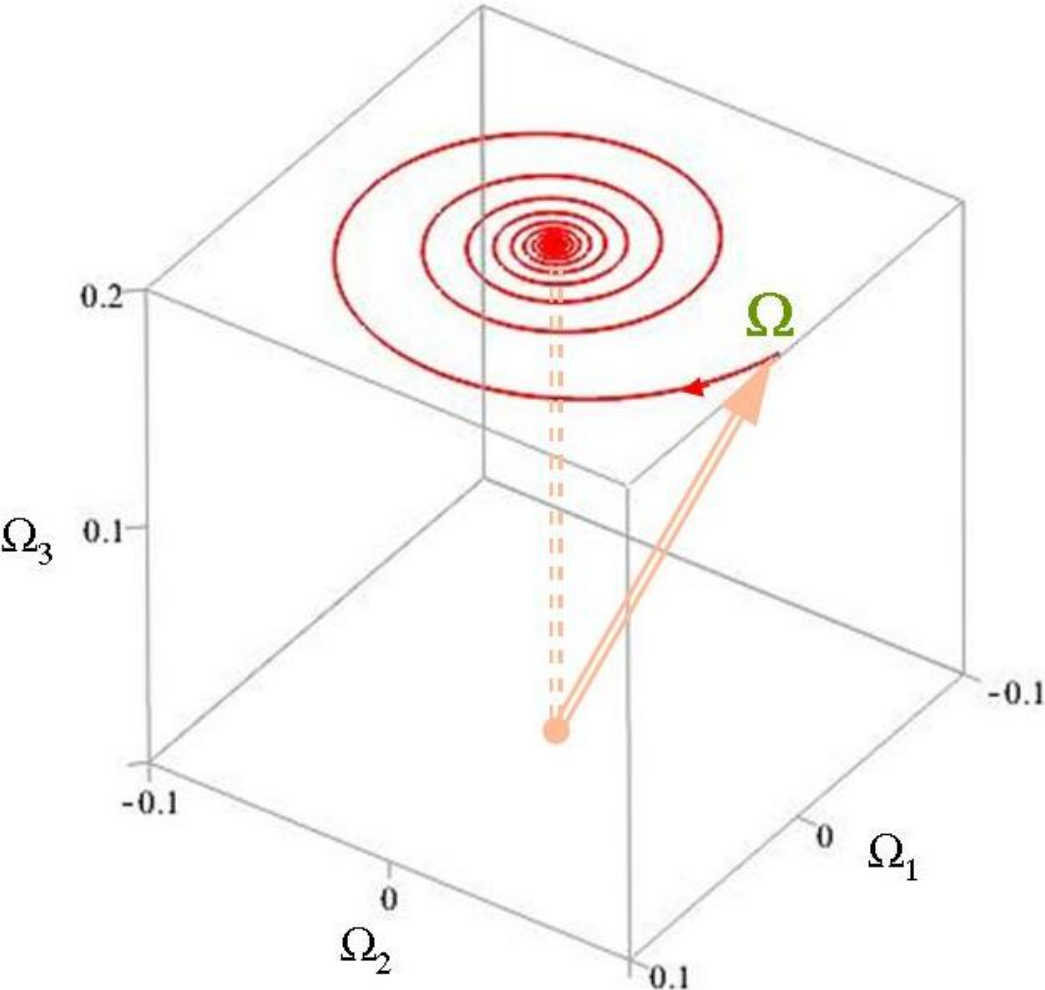}\hspace{4ex}
\includegraphics[scale=0.7]{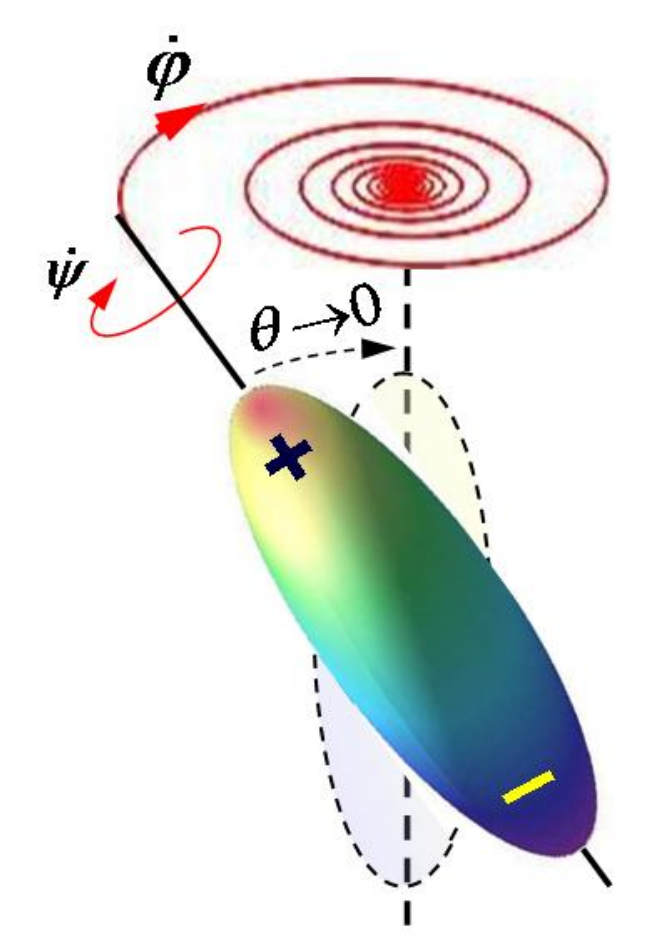}
\vspace{-2ex} \caption{The same as in figure 1, but according to the reduced
ALE equations: the inclination angle $\theta$ in asymptotics
$t\to\infty$ tends exponentially to zero while the proper rotation speed
tends to the constant $\Omega_3$.}
\end{center}
\end{figure}
%------------------------------------------------------------

According to \cite[\S\,75]{L-L87E}, the equation \re{2.2} is derived
from the expression for the force of the dipole radiation reaction,
i.e., the Abraham-Lorentz expression generalized for a system of
several charges.

Since the Abraham-Lorentz equation is generally accepted, the
balance equation \re{2.2} should be preferred. Let us note that the
authors of the textbook \cite{L-L87E} apparently considered this
equation as an intermediate formula (unnumbered in
\cite[\S\,75]{L-L87E}), which is further reduced to the previously
obtained (Problem 2 in \cite[\S\,72]{L-L87E}) equation \re{2.1} by
neglecting the Schott term. It was noted that this step is
admissible provided the motion is stationary (but in
\cite[\S\,75]{L-L87E} the nearly stationary motion apparently
implied). In the present case a free symmetric Euler spinning-top
makes certainly a stationary motion but the taking the radiation
reaction into account without the Schott term and with it leads to a
nearly stationary motions with quite different final states of the
top. Once the balance equation \re{2.2} is plausible, then there is
something wrong with the equation \re{2.1} or its usage.

Similarly to the Larmor formula, the balance equation \re{2.1} was
derived in \cite{L-L87E} (\S\,72, Problem 2) by means of an
integration of the angular momentum flux over a sphere of some
radius $R_0$ surrounding a system of charges.

According to one common interpretation \cite{Gro11}, the Schott
energy (and the Schott angular momentum in our instance) of
nonrelativistic system (which is the case) is not present in the
Larmor formula (here -- in eq. \re{2.1}) because it is localized
somewhere close to charges, i.e., deep under the integration sphere.

But the neglect of the Schott term leads in our case to an
irreparable ``loss'' of the remnant energy $E=I_3\Omega_3^2/2$ and
the angular momentum $ L_3=I_3\Omega_3$ via radiation across the
integration surface. Thus this interpretation does not clarify the
current problem.

According to another, later interpretation by A.~Singal \cite{Sin16c},
there is no need to attribute spatial localization and even physical
essence to the Schott energy. When considering the Larmor formula
one usually does not take into account the fact that although the
radius of the integration sphere $R_0$ falls out of the final
formula (see, for example, (67.8) or (67.9) in \cite{L-L87E}), its
right-hand side refers to the moment of time delayed by $ R_0/c $
that is the value required to reach by electromagnetic signal the
sphere from its center.

The recalculation of the Larmor formula to ``real'' time, carried
out with the spherical-shell model of a charged particle
\cite{Sin16c}, reproduces the Schott term in this formula, and thus
removes the inconsistency of the energy balance equations. The
recalculation, however, refers to the electromagnetic structure of
charged particles, and hence that structure may be considered as a
carrier of the acceleration-dependent Schott energy. This interpretation
following from Singal's calculations disagrees
with his own interpretation but is consistent with the conception of
charged particles as extended, composite, and therefore non-rigid
objects due to their electromagnetic structure \cite{Row10}.

In the Appendix, a similar recalculation is performed in the angular momentum
balance equation of a charged particle. By this it is approved the correctness
of the formula \re{2.2} and the incorrect application of the formula \re{2.1}.

%%%%%%%%%%%%%%%%%%%%%%%%%%%% APPENDIX %%%%%%%%%%%%%%%%%%%%%%%%%%%%%%%%%

\section*{Appendix. On interpretation of the Shott term in the angular momentum balance equation.}
\renewcommand{\theequation}{A.\arabic{equation}}
\setcounter{equation}{0}

Following  the Singal's calculations \cite{Sin16c} (but with different
interpretation in mind), we present the charged particle as a sphere
of small radius $r_0$, and its mass as the sum of the ``bare'' mass
$m_0$ and the electromagnetic mass $m_{\rm em}=2q^2/(3c^2r_0)$. Then
the equation of motion can be represented in such a way
%
%           Equation A.1
\begin{equation}\lab{A.1}
\frac{\D{}{}}{\D{}t}\big\{m_0{\B v}(t)+m_{\rm em}{\B
v}(t-r_0/c)\big\}+O(r_0)=\B F_{\!\!\rm ex}(t),
\end{equation}
that in the limit $r_0\to0$ the total mass $m=m_0+m_{\rm em}$ keeps
finite, and the equation \re{A.1} goes into \re{2.5}. The 1st term
in curly braces represents a ``bare'' contribution in the particle
momentum, the 2nd one is a contribution of its electromagnetic
``fur''; the later perceives a change of position, velocity etc.
with delay because of a finite size of the particle.

Hereafter the moment of time $t$ as the argument of physical
quantities will be omitted, and the retarded moment of time
$t-r_0/c$ will be denoted by index ``ret''.

Let us multiply vectorially $\B r$ by equation \re{A.1},
use the Taylor series expansion up to the 1st order,
$\B r\simeq[\B r]_{\rm ret}+[\dot{\B r}]_{\rm ret}r_0/c$,
and rearrange terms:
%
%           Equation A.2
\begin{equation}\lab{A.2}
\frac{\D{}{}}{\D{}t}\big\{m_0{\B r}\times{\B v}+m_{\rm em}[{\B
r}\times{\B v}]_{\rm ret}\big\}+O(r_0)=-\frac{2q^2}{3c^3}[{\B
v}\times\dot{\B v}]_{\rm ret}.
\end{equation}
This equation clears up a meaning of the equation \re{2.8}:
on the left in curly braces is a total angular momentum of the particle
consisting of  ``bare'' and electromagnetic contributions. On the right
is a result of the integration of the angular momentum flux over the sphere
which radius cannot be smaller than $r_0$ (hence the delayed argument).
In the limit $r_0\to0$ we have
%
%           Equation A.3
\begin{equation}\lab{A.3}
\lim_{r_0\to0}m_{\rm em}\big\{[{\B r}\times{\B v}]_{\rm ret}-{\B
r}\times{\B v}\big\}=-\tau_0\dot{\B L}\equiv-{\B L}_{\rm S},
\end{equation}
where $\tau_0=2q^2/(3mc^3)$. This reduces the equation \re{A.2} to \re{2.6}.
Obviously, one can treat in a similar manner the relation of the balance
equations \re{2.1} and \re{2.2} for a system of charges.

\bigskip

\providecommand{\newblock}{}

\end{document}